\documentclass[twoside,twocolumn,english,10pt]{article}
\usepackage[dvips]{graphicx}
\usepackage{rsc}
\usepackage{latexsym}
\usepackage{graphicx}
\usepackage[export]{adjustbox}
\usepackage{float}
\usepackage[utf8]{inputenc}
\usepackage[english]{babel}
\usepackage{amsmath}
\usepackage{mathtools}
\usepackage{amssymb}
\usepackage{verbatim}
\usepackage{amsfonts}
\usepackage{hyperref}
\usepackage{authblk}

\usepackage{enumitem}
\usepackage{color}

\definecolor{blue}{rgb}{0,0,1}
\definecolor{green}{rgb}{0,.6,0}
\definecolor{red}{rgb}{1,0,0}
\definecolor{vio}{rgb}{1,0,1}
\definecolor{uv}{rgb}{0.5,0,0.5}
\definecolor{ama}{rgb}{0.3,0.3,0.3}

  \usepackage{verbatim}
  \newcommand{\dif}{\mathrm{d}}

  \newcommand{\Z}{\mathbb{Z}}
   \newcommand{\im}{\mathbf{i}}
    
   \newcommand{\abs}[1]{\left\vert#1\right\vert}

    \setlist[itemize]{leftmargin=*} 

\usepackage{sectsty}\usepackage{balance}

\usepackage{lastpage}\usepackage[format=plain,justification=raggedright,singlelinecheck=false,font=small,labelfont=bf,labelsep=space]{caption}\usepackage{fancyhdr}

\begin{document}

\title{ Soft modes and strain redistribution in continuous models of amorphous plasticity: the Eshelby paradigm, and beyond?}
\author[1,2]{Xiangyu Cao} \author[1]{Alexandre Nicolas}\author[1,3]{Denny Trimcev}   \author[1]{Alberto Rosso}
\affil[1]{LPTMS, CNRS, Univ. Paris-Sud, Universit\'e Paris Saclay, Orsay, France}
\affil[2]{Department of Physics, University of California, Berkeley, Berkeley CA 94720, USA}
\affil[3]{Department of Physics, École Normale Supérieure, F-75231 Paris Cedex 05, France}
\date{\today}

\twocolumn
[
	\begin{@twocolumnfalse}

	\maketitle

	\begin{abstract}
	
The deformation of disordered solids relies on swift and localised rearrangements of particles. The inspection of soft vibrational modes can help
predict the locations of these rearrangements, while the strain that they actually redistribute mediates collective effects.
Here, we study soft modes and strain redistribution in a two-dimensional continuous mesoscopic model based on a Ginzburg-Landau free energy for perfect solids,
supplemented with a plastic disorder potential that accounts for shear softening and rearrangements. 
Regardless of the disorder strength, our numerical simulations show soft
modes that are always sharply peaked at the softest point of the material (unlike what happens for the depinning of an elastic interface). Contrary to
widespread views, the deformation halo around this peak does not always have a quadrupolar (Eshelby-like) shape. Instead, for finite and narrowly-distributed disorder, it looks like a fracture, with a strain field that concentrates along some easy directions. These findings are rationalised 
with analytical calculations in the case where the plastic disorder is confined to a point-like `impurity'. In this case, we unveil a continuous
family of elastic propagators, which are identical for the soft modes and for the equilibrium configurations. This family interpolates between the standard quadrupolar propagator and the fracture-like one as the anisotropy of the elastic medium is increased. Therefore, we expect to see a fracture-like
propagator when extended regions on the brink of failure have already softened along the shear direction and thus rendered the material
anisotropic, but not failed yet. We speculate that this 
might be the case in carefully aged glasses just before macroscopic failure.
	\end{abstract}

  \end{@twocolumnfalse}

\section{Introduction}
]
\begin{figure}[h]
	\begin{center}
		\includegraphics[width=0.8\columnwidth]{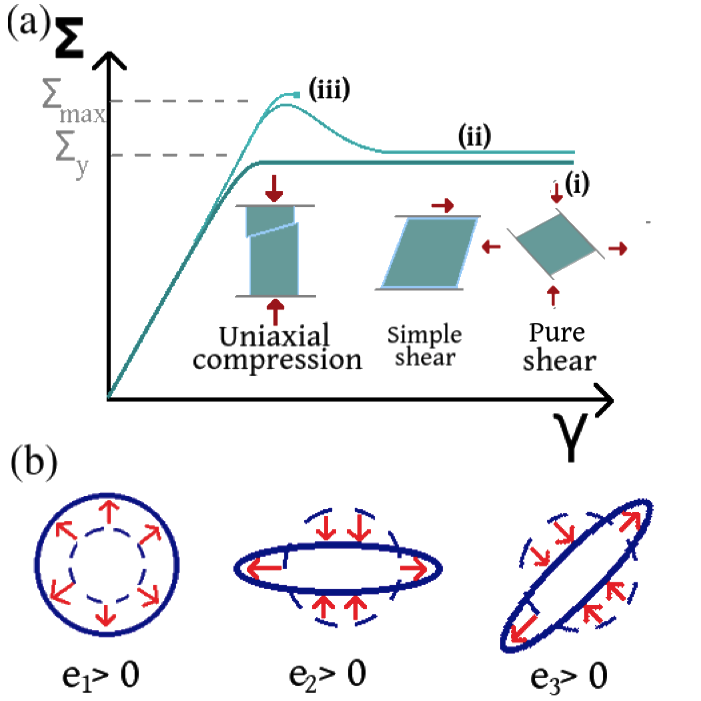}
	\end{center}
	\caption{(a) Sketch of the macroscopic shear stress response of a disordered solid subject to a quasistatic deformation,
		with depictions of common deformation protocols. Stress fluctuations are not represented in the sketch. 
		(b) Representations of the new strain variables $e_1$, $e_2$, and $e_3$. In this work we consider pure shear along the $e_2$ direction and $\gamma$ identifies to the average of $e_2$.}\label{fig:Macroscopic_response} 
\end{figure}
Apply a dab of toothpaste onto a toothbrush and slightly tilt the brush. The paste will respond to the small shear stresses $\Sigma$ 
thus created in its bulk by deforming elastically. In contrast, when you squeeze a toothpaste tube, the stresses in the material exceed a critical yield value $\Sigma_y$, and the paste starts to flow. This ``liquid''-like phase under shear is observed not only in pastes and (concentrated) suspensions,
 but also in other soft solids such as emulsions and foams \cite{coussot2014yield}. Other disordered materials such as metallic
glasses also depart from an elastic behavior under large enough stresses, but then break instead of flowing. In the athermal limit, the change observed at $\Sigma_y$ is a dynamical phase transition known as yielding transition.

To study it, a standard experimental protocol consists in slowly\footnote{One may regard the deformation as slow if
the stress-strain curve does not substantially vary when the driving rate is reduced.} deforming the material and monitoring its macroscopic stress. For small deformations, the response is linear and elastic. For larger ones, the deformation becomes macroscopically irreversible, due to the onset of plasticity. Three distinct 
plastic responses can be observed, as shown in Fig.~\ref{fig:Macroscopic_response}:
(i) the stress grows monotonically and saturates at a steady-state value $\Sigma_y$; (ii) the stress overshoots $\Sigma_y$ and, upon reaching $\Sigma_\text{max}>\Sigma_y$, drops rapidly to the stationary value $\Sigma_y$ \cite{divoux2011stress} (note that it is still unclear if
 the material fails globally at $\Sigma_\text{max}$ -- as in a spinodal transition  \cite{zapperi1997first,procaccia2017mechanical} -- or through a large sequence of finite-size avalanches) (iii) at $\Sigma_\text{max}$, the material breaks and the stress drops to zero.

Microscopically, the mechanism underlying the irreversible plastic response is the localised rearrangement of a few particles
 (droplets in emulsions, bubbles in foams), a process called shear transformation (ST)~\cite{argon1979plastic}. Recently, more detailed investigation has revealed that these ST do not arise randomly in the material, but display spatial correlations \cite{chikkadi2012shear,nicolas2014spatiotemporal}. It is now widely believed that
 these correlations stem from the elastic deformation induced by the ST, which has a peculiar quadrupolar
 shape, as predicted by Eshelby half a century ago \cite{eshelby1957determination, schall2007structural}. This quadrupolar kernel has been observed in atomistic simulations of several model glasses \cite{maloney2006amorphous,puosi2014time}. Moreover, the  plastic ST instability is preceded by the emergence of localisation in the low-frequency modes of the vibrational spectrum~\cite{tanguy10mode,manning11soft,charbonneau16universal}. The localised soft spots tend to coincide with the subsequent ST. Interestingly, the displacement field around a soft spot displays a long-range tail, decaying as $r^{1-d}$, with $d$ the spatial dimension, which is  consistent with the quadrupolar shape observed after the plastic event.     

On the basis of this picture of localised plastic rearrangements, elasto--plastic models (EPM) have proposed to coarse-grain disordered solids into
a collection of blocks alternating between an elastic regime and plastic events interacting via a quadrupolar kernel. Following
similar endeavours for the study of earthquakes \cite{chen1991self}, these models have succeeded in capturing the presence of strongly correlated
dynamics in these systems (avalanches, possible shear bands, etc.)~\cite{baret2002extremal,budrikis2013avalanche,nicolas2017deformation, lin2014scaling, lin15prl,gueudre17scaling}. However, a clear connection between the microscopic description and these coarse-grained models is missing. In particular, the
universality of the quadrupolar propagator used in EPM may still be questioned and the discreteness of EPM precludes the study of vibrational modes.

An alternative approach is provided by continuum models that extend the free energy description of solids beyond the perfect elastic limit. In these models plasticity is introduced by means of a disordered potential which displays many local minima, as explained in  Section~\ref{sec:model}. 
Such models, possibly pioneered by Kartha and co-workers~\cite{kartha95tweed}, were intensively studied by Onuki~\cite{onuki2003plastic} and Jagla~\cite{jagla07shear}. This paper intends to use the continuum approach to bridge the gap between atomistic simulations and discrete EPM
\footnote{See \cite{salman2011minimal} for a related endeavour to connect continuous models with discrete automata in the context of crystal plasticity.}, with an emphasis on the initial soft modes and the actual response to ST.

\begin{figure*}
	\center
\includegraphics[width=.9\textwidth]{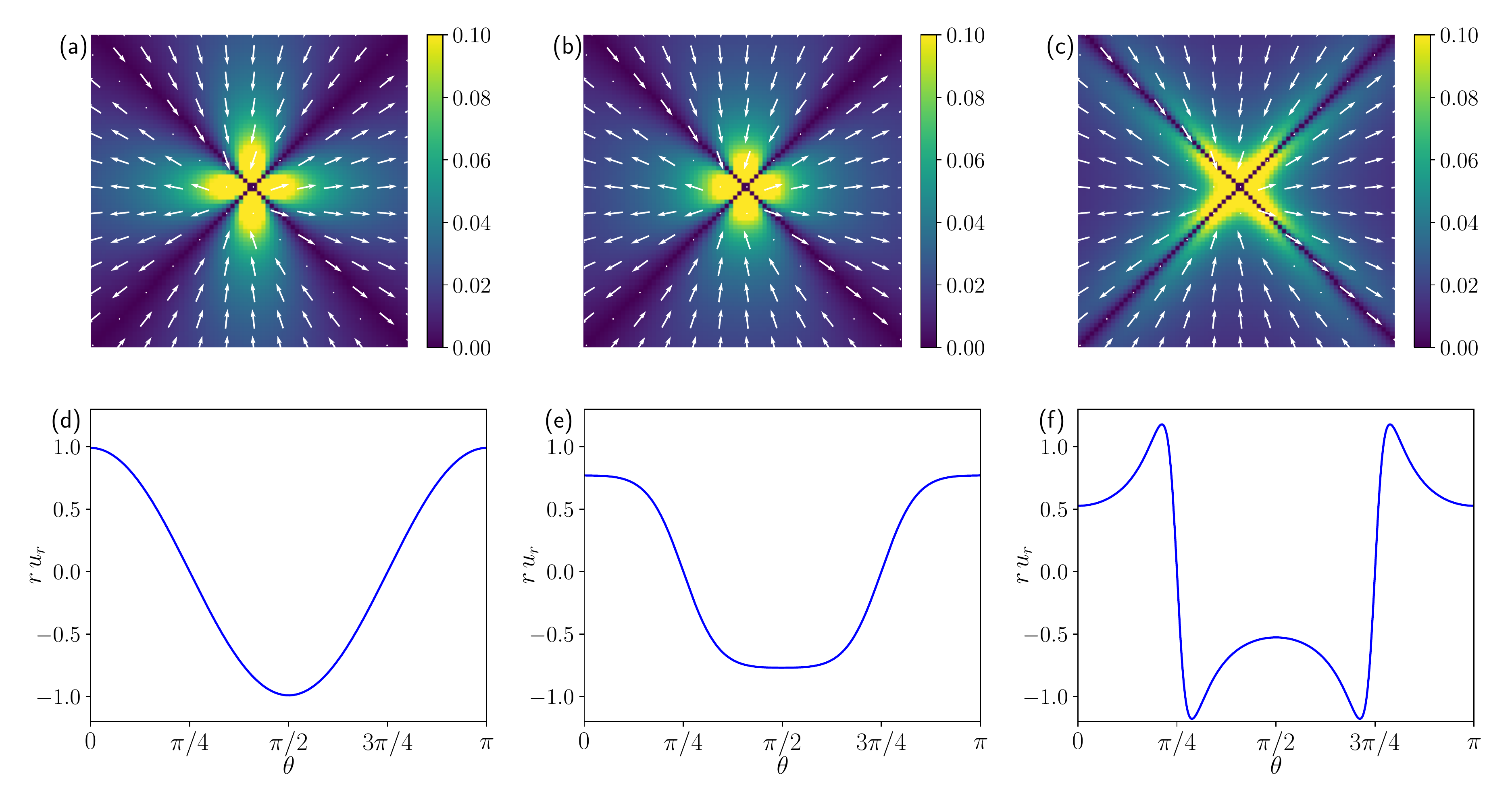}
\caption{Steady-state response to a ST located at the origin and embedded in an elastic medium with $\delta=$ 0.01 (a,d), 0.3 (b,e) and 0.9 (c,f). The top row shows the displacement field. The white arrows represent the displacement fields $u$ while the colour code denotes the displacement amplitude $\left\Vert u \right\Vert$. The bottom row shows the azimuthal variations of the radial displacement multiplied by $r$. (a,d) corresponds to the standard quadrupolar kernel, while (c,f) represents a fracture-like profile, in which the displacement is maximal along the directions of macroscopic failure $\theta = \pm \pi/4, \pm 3\pi/4$.  }\label{fig:displacement_th} 
\end{figure*}

Considering two-dimensional (2D) materials subjected to pure shear, we find that the low-frequency modes are always peaked in point-like  ``soft spots'', where the next ST will take place. This extreme localisation is at variance with short-range depinning models, where soft modes have a finite localisation length which can be tuned by playing with the disorder strength~\cite{cao2017localisation,Tanguy2004localisation}. A closer analysis shows a halo of finite displacements around the soft spots pointing in the radial direction, with a $1/r$ radial decay and a two-fold azimuthal symmetry (this corresponds to a $1/r^2$ decay with four-fold azimuthal symmetry in the strain field), due to the elastic embedding of the impurities, see Section~\ref{sec:denny}. Surprisingly, these halos do not always match Eshelby's solution. Instead, we find a one-parameter continuous family of kernels depending on the distribution of plastic disorder in the system (see Fig.~\ref{fig:displacement_th}). Their shapes are rationalised analytically in Section~\ref{sec:single_impurity} by calculating the soft mode associated with a point-like plastic impurity at $r=0$ embedded in an 
incompressible elastic medium. In polar coordinates, the 
(non-affine) displacement field $u$ reads:
\begin{equation}
u_r(r,\theta) \propto \frac{ \cos(2\theta)} {1 + \delta \cos(4\theta)} \,,\, u_\theta = 0\,, \label{eq:u}
\end{equation}
where $\theta = 0$ is the principal axis of positive stretch, and $\delta=(\mu_3-\mu_2)/(\mu_3+\mu_2)$ quantifies the plasticity-induced anisotropy in the shear moduli $\mu_2$ and $\mu_3$ (associated with the strains $e_2$ and $e_3$, respectively; see Fig~\ref{fig:Macroscopic_response}).  For $\delta=0$ we recover the standard quadrupolar (Eshelby-like) propagator. When $\delta \to 1$ we find
a fracture-like kernel concentrating the deformation along the diagonal directions. This limit is obtained when the plastic potential softens the material to such an extent that the modulus along these directions vanishes (namely $\mu_2 \to 0$ while $\mu_3$ remains finite). To the best of our knowledge, the fracture-like propagator has not been observed yet, but we speculate that it might be seen in carefully aged glasses, in the marginal state that precedes global failure, when extended regions are on the brink of plastic failure.
In the case of a single impurity, the soft mode has exactly the same shape as the final strain field induced by the ST,  but also closely
(or exactly if $\mu_2=\mu_3$)
matches the transient strain field during the plastic event, up to renormalisation (Section~\ref{sec:SPE3}).

\section{Field-based models \label{sec:model}}
To begin with, we recall how plasticity is introduced in Continuum Mechanics descriptions of disordered solids \cite{kartha95tweed, puglisi2000mechanics,onuki2003plastic, lookman03ferro,jagla07shear,jagla2017non}.
In the spirit of the works of Jagla \cite{jagla07shear}, this is achieved by first writing the free energy of an elastic material and then incorporating the plastic disorder in it.

\subsection{Strain variables and linear elasticity}
Even though glassy materials are discrete at the atomic scale, they can be handled as continua as long 
as one is interested in length scales larger than a few particle diameters \cite{tsamados2009local}.
To linear order, deformations in a continuous medium are quantified by the strain tensor
\begin{align}
&\epsilon_{ij} = \frac12(\partial_j u_{i} +  \partial_i u_{j}) \text{ with } i,j= 1,2\text{ in 2D.}  \label{eq:tensors}
\end{align}
It is convenient to trade off the strain tensor $\epsilon_{ij}$  for the following three strain variables: the volume distortion $e_1 = (\epsilon_{11} + \epsilon_{22})/2$ and the two independent shear strains 
$e_2 = (\epsilon_{11} - \epsilon_{22})/2$ and $e_3 =  \epsilon_{12}$, represented in Fig.~\ref{fig:Macroscopic_response}(b).

Since the $e_i$ derive from the same displacement field, the commutation rule
for partial derivatives, \emph{i.e.}, $\partial_{jk} u_{i} = \partial_{kj} u_{i}$,
imposes a constraint on these variables, known as St. Venant condition. In terms of the Fourier transforms 
$ \hat{e}_i (q) = \int e_i(r) e^{\im q.r}  \dif^2 r $, this constraint reads
\begin{align}
&g:=\sum_{i=1}^3 Q_i \hat{e}_i = 0 \,,\, \label{eq:StV}  \\
&(Q_i)_{i=1}^3 = (-\abs{q}^2, q_1^2 - q_2^2, 2 q_1 q_2) \,, \label{eq:Ds} 
\end{align}
where $q_j = -\im \partial_j$ are the differential operators in Fourier space. These conditions turn out to be sufficient to prove the existence of the displacement field. 

With these new variables, the free energy of a uniform linear elastic solid reads 
\begin{equation}
F = \int_r B e_1^2 + \mu_2 e_2^2 + \mu_3 e_3^2 \label{eq:elastic_free} \,,
\end{equation}
where $B$ is the bulk modulus and $\mu_2$ and $\mu_3$ are the
shear moduli. The incompressible limit corresponds to  $B\rightarrow \infty$ while, for an isotropic material, $\mu_2=\mu_3$.

\subsection{Plasticity}
If a disordered solid is strongly sheared,
it will start to respond plastically by accumulating 
irretrievable deformation. This irreversible response is accounted for by introducing an
anharmonic contribution to the free energy $F$ of eq.~\ref{eq:elastic_free} so that $F$ can have multiple local minima, viz.,
\begin{equation}
F = \int_r \left[B e_1(r)^2 + V_{2,r}[e_2(r)] + \mu_3 e_3(r)^2 \right] dr \label{eq:free} \,,
\end{equation}
where $V_{2,r}(e_2)= \mu_2 e_2^2 + W_{r}(e_2)$, with typically $ W_{r}(e_2)\leqslant 0$. Here, and in all the following,
we will consider that plasticity only develops along the macroscopic shear direction, chosen to be $e_2$.  Schematically, when the local strain $e_2$ at, say, $r=0$ is driven to a local maximum of $V_{2,r=0}$, a plastic event begins in which $e_2(0)$ slides into the next potential valley. This local change
is elastically coupled to other regions of the material and may trigger other plastic events at $r^\prime \neq 0$, if $V_{j,r^\prime}(e_j)$ is also anharmonic. 

It is interesting to remark that, because of the shear-softening contribution $W_r$ to $V_{2,r}$, the local tangential shear modulus $\tilde{\mu}_2(0)= \frac{1}{2} V_{2,r=0}''$ vanishes at the onset of the instability,
as observed in atomistic simulations. The softening of $\tilde{\mu}_2$ has the same effect
on the local stress-strain relation 
as a shift of the minimum of the potential $V_{2,r=0}$ by an amount $e^{pl}$ in a material
where the shear modulus would remain constant ($\tilde{\mu}_2=\mu_2$), \emph{viz.}, 
\begin{align}
&\sigma(r) = V_{2,r}^\prime[e_2(r)] = 2 \mu_2 [ e_2(r) -  e^{pl} ]\,,\, \\
 \text{where } & e^{pl} := - \frac{1}{2\mu_2}W_r^\prime[e_2(r)] \,. \label{eq:epldef}
\end{align}

\begin{figure}[t]
	\center
	\includegraphics[width=.7\columnwidth]{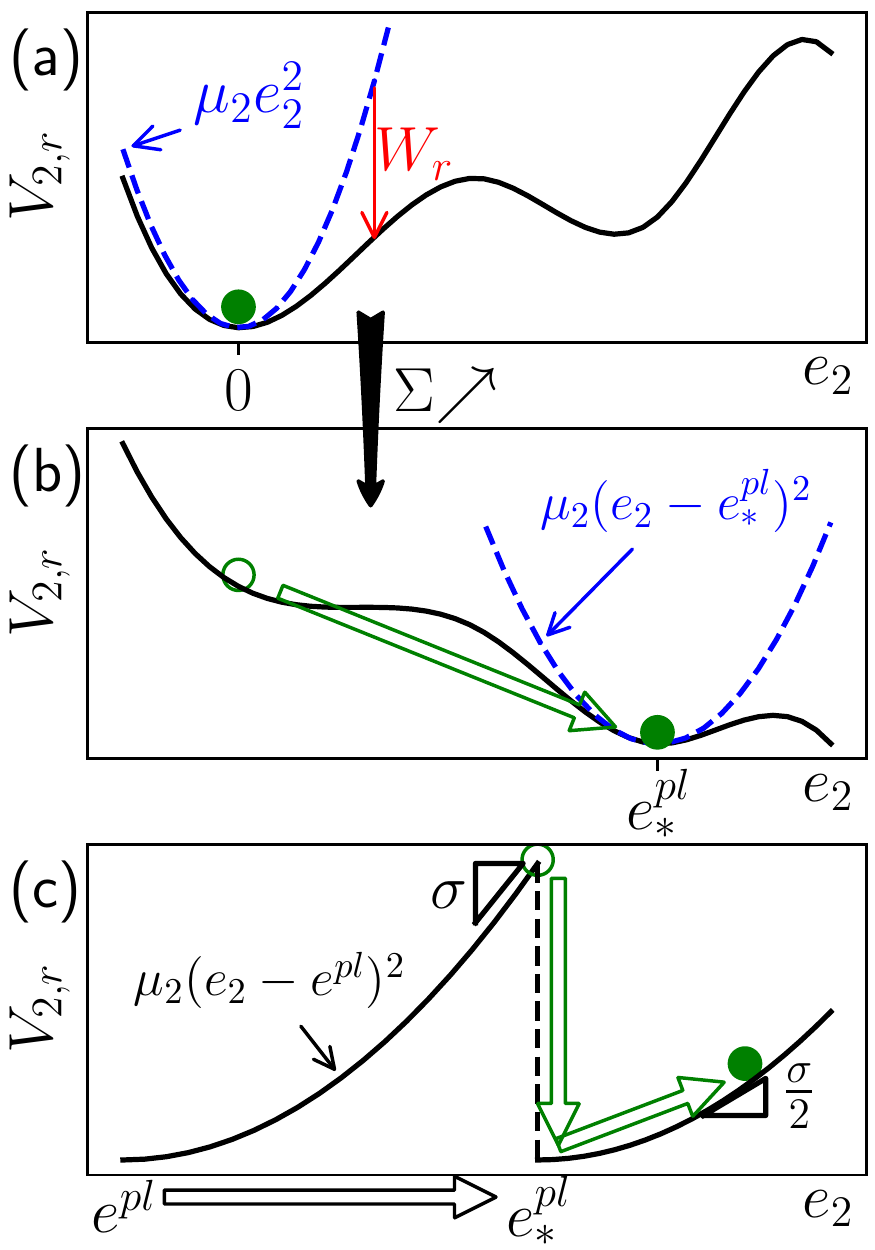}
	\caption{ \emph{Illustrations of possible disordered potentials $V_{2,r}(e_2)$ --} (a,b) Interpretation of a plastic
		rearrangement from the viewpoint of the potential: While the system sits in a linear elastic region (a), the 
		applied macroscopic stress $\Sigma$ tilts the potential, so much so that the system starts exploring the shear-softening
		region ($W_r<0$) and is finally able to briskly slide into a deeper minimum (b). 
		(c) Illustration of the disordered potential used to derive rules of elasto-plastic models. The local stress variations during a plastic event are highlighted by green arrows and explained in the main text. }\label{fig:V2} 
\end{figure}

\subsection{Dynamics}
Turning to the dynamics, the evolution of $e_i$ in the overdamped limit (relevant for foams and concentrated emulsions) is obtained by differentiating the total free energy $F_\mathrm{tot}$, viz. 
\begin{equation}
2 \eta_i \dot{e}_i = - \frac{\delta F_{\text{tot}}}{\delta e_i} \,,\label{eq:EoM} 
\end{equation}
where $\eta_i$ refers to the effective microscopic viscosity associated with the strain $e_i$. A difference with respect to the equation describing the depinning of 
an elastic line deserves to be underscored here. In the depinning case, the damping acts on the velocities $\dot{u}_i$, due to friction against a fixed
substrate (or medium). Here, there is no such substrate and dissipation is due to the non-uniform velocities within the system; thus, only the velocity 
gradients are damped, as in the Navier-Stokes equation for fluids. 

In addition to the terms in eq. \ref{eq:free}, the total free energy $F_{\text{tot}}$,  includes two contributions: the driving along $e_2$, $- \Sigma \,  \hat{e}_2(0)$, where $ \Sigma$ corresponds to the macroscopic stress, as well as the St Venant constraint (eq.~\ref{eq:StV}) for all $q\neq0$ enforced by the Lagrange multipliers $\lambda(q)$, viz.,
\begin{equation}
F_{\text{tot}} = F + \int_{q \neq 0} \hat{\lambda}(q) \hat{g}(q)  - \Sigma\, \hat{e}_2(0) \,. \label{Ftot}
\end{equation}
 The St. Venant constraint is naturally obtained by extremising $F_\text{tot}$ with respect to $\lambda(q)$, i.e., solving $\hat{g}(q) = 0$. 

The system can be driven in two distinct ways: \emph{strain-controlled} protocols consist in controlling the time evolution of the macroscopic strain $\hat{e}_2(0)$ by properly adjusting $\Sigma$. Conversely, \emph{stress-controlled} protocols result from imposing a constant macroscopic stress $\Sigma$ in eq.~\ref{Ftot} and leaving $\hat{e}_2(0)$ free. Via eqs.~\eqref{eq:EoM}-\eqref{Ftot}, the macroscopic stress can be expressed as $\Sigma= \frac{1}{V} \int_r \sigma(r) + \eta_2 \dot \gamma$, where $\dot{\gamma}= \frac{2}{V} \dot{\hat{e}}_2(0) $ is the shear rate. If $\Sigma$ exceeds a value $\Sigma_y$ (set by the potentials
$V_{2,r}$), the system will flow forever, with an ever--increasing strain $\hat{e}_2(0) \sim \dot{\gamma} t$ in the stationary state. Otherwise, it will reach a new equilibrium.



\subsection{Incompressible limit}
From now on, we will focus on the incompressible limit $B\rightarrow \infty$, but equations for compressible systems are 
provided in Appendix~\ref{sec:compressible}.  In this limit, $e_1=0$ so the St. Venant constraint yields 
\begin{equation} 
2 q_1 q_2 \hat{e}_3(q) = (q_2^2 - q_1^2) \hat{e}_2(q) \,. \label{eq:StV-incomp}
 \end{equation}
As a consequence,  the Fourier modes $\hat{e}_2(q)$ vanish whenever $q_1=0$ or $q_2=0$, with $\abs{q}>0$. In real space, this means that the integral of $e_2$ along each horizontal or vertical line is constant and equal to the average strain, regardless of the mechanical law. Similarly, the integral of $e_3$ along each $\pm \pi/4$-direction line is also constant. These constraints on the strains are more fundamental than their counterparts for the stresses (or elastic strains), which are largely used in elasto-plastic models. The latter constraints impose that $\sigma_2$ ($\sigma_3$) have constant average along $\pm \pi/4$ lines (horizontal/vertical lines, respectively), where the stress components ${\sigma}_j$'s are in direct correspondence to the strain variables $e_j$, but these constraints are derived under the assumption of mechanical equilibrium.


%

In addition, the displacement field $u_i$, which is easier to interpret than $e_2$, reduces to:
\begin{equation}
\hat{u}_1  = \im q_1^{-1} \hat{e}_2 \,,\, \hat{u}_2 = -\im q_2^{-1} \hat{e}_2 \, \label{eq:integration}
\end{equation}
if $q_1 q_2\neq 0$ and $\hat{u}_1,\hat{u}_2=0$ otherwise. Note that the zero mode $q=(0,0)$ of $\hat{u}_j$ corresponds to a global translation of the system.
Hereafter, we assume isotropic viscosities, which are set to unity, viz., $\eta_2= \eta_3= 1$. Thus, the evolution of $e_2$, eq.~\ref{eq:EoM}, turns into
\begin{align}\label{eq:motion}
2 \dot{e}_2 = & - \frac{\delta F}{\delta e_2} + Q_2 \frac{ Q_2 \frac{\delta F}{\delta e_2} + Q_3 \frac{\delta F}{\delta e_3}}{ Q_2^2 + Q_3^2} + \Sigma 
\end{align}
and simplifies to
\begin{equation}
2 \dot{e}_2  =  -2\mu_2 e_2 -W_r^\prime(e_2) + \Sigma  - \mathcal{G}_q [ W_r^\prime(e_2) - 2 \delta \mu e_2 ] ,  \label{eq:EoM_inc_real}
\end{equation}
where we have used the shorthands $\delta \mu =\mu_3 - \mu_2$ and
\begin{equation} \mathcal{G}_q=\frac{-(q_1^2 - q_2^2)^2}{q^4} \label{eq:Gq} \end{equation} 
for $q\neq 0$, and $\mathcal{G}_{q=0} = 0$. For systems with finite inertia, the foregoing overdamped equation can be generalised:
  \begin{align}
  2 \rho  \ddot{e}_2 = -q^2 \left[ \frac{1+\mathcal{G}_q}2 W_r^\prime(e_2)   +  \mu_2 e_2  -  \delta \mu \, \mathcal{G}_q  e_2  +  \dot{e}_2 \right]  \,,
     \end{align}
where $\rho$ is the density (see Appendix~\ref{app:underdamped}).

Mathematically, one may notice that, for any given $q \neq 0$, the incompressible St Venant constraint~\eqref{eq:StV-incomp} is satisfied on
the line $e_3= -Q_2 Q_3^{-1} e_2$ of the plane $(\hat{e}_2(q),\hat{e}_3(q))$ and that 
the action of the Lagrange multiplier $\hat{\lambda}(q)$ is equivalent to taking only the tangential component of
$\nabla F:= \left[ \delta F / \delta \hat{e}_2, \delta F / \delta \hat{e}_3 \right] $ along this line,
via a projection of $\nabla F$ onto the \emph{unit} tangential vector $\boldsymbol{t} := (Q_3/q^2,-Q_2/q^2)$ [note that $Q_2^2 + Q_3^2 = q^4$  through eq.~\eqref{eq:Ds}]. Therefore,
it is incorrect to just express $F$ as a function of $e_2$ on this line, \emph{viz.}, $\tilde{F}(e_2) := F(e_2, -Q_2 Q_3^{-1} e_2)$
and then omit the Lagrange multiplier. Indeed, differentiating $\tilde{F}$ with respect to $e_2$ is equivalent to projecting 
$\nabla F$ onto the \emph{non-normalised} tangential vector $\boldsymbol{\tilde{t}} = \left[ 1,-Q_2 Q_3^{-1} e_2\right]$.
To obtain the correct dynamics without using multipliers, one should parametrise $F$ with the properly rescaled tangential coordinate
\begin{equation}
\hat{e} (q):=  \left[ \hat{e}_2(q), \hat{e}_3(q) \right] \cdot \boldsymbol{t}  = q^2 Q_3^{-1} \hat{e}_2 (q) \,, \label{eq:rescaled_strain} 
\end{equation}
where $\cdot$ denotes the inner product.The rescaled coordinate $\hat{e}$ will prove useful in Sec.~\ref{sec:denny}.

\section{Response to a single plastic event \label{sec:single_impurity}}
Although the equations established in the previous section fully define the model once the disorder potential $W_r$ is chosen,
solving the full problem is complex. It is thus enlightening to start by considering the simple case where the plastic disorder
is confined to an `impurity' of size $a \to 0$ around $r=0$, while the rest of the material is elastic, 
 viz., 
\begin{align}
V_{2,r}(e_2) =& \mu_2 e_2^2(r) + a^2 \delta_{2D}(r) W(e_2(0)) - \Sigma\, \hat{e}_2(0) \,, \label{eq:V2oneimp}
\end{align} %
where $W$ is a disordered potential and $\delta_{2D}(r)$ is the Dirac distribution in 2D, evaluated at position $r$. Note that, for convenience, we have incorporated the driving contribution $- \Sigma\, \hat{e}_2(0)$
to the total free energy $F_\mathrm{tot}$ of eq.~\eqref{Ftot} into $V_{2,r}$. Beginning with the stable configuration
sketched in Fig.~\ref{fig:V2}(a), the driving gradually tilts the potential until the configuration becomes unstable 
(Fig.~\ref{fig:V2}(b)). This triggers an instability whereby the local strain $e_2(0,t=0)$ evolves rapidly with time $t$
and modifies the local plastic disorder $W^\prime[e_2(0,t)]$ and plastic strain $e^{pl}(t)=- W^\prime[e_2(0,t)] / 2\mu_2$.
Finally, the next stable configuration is attained, after a plastic strain $e^{pl}= e^{pl}_\star$ has been cumulated.

In elasto--plastic models, the mechanical equilibration time is neglected everywhere in the material, except in the plastic
impurity \cite{nicolas2017deformation}; thus, the strain field $e_2$ is always an equilibrium configuration for a given plastic
strain $e^{pl}$, but this plastic strain may need a finite time to reach its final value $e^{pl}_\star$. In this section,
we first derive the equilibrium configurations for the continuous models under study here and then show that, up to
a normalisation coefficient, they coincide with the soft modes of the system just before the onset of the instability. Finally,
we explore the transient dynamics during the plastic event.

During the whole evolution, the dynamics are governed by the equation of motion~\eqref{eq:EoM_inc_real}. For Fourier modes $q \neq 0$ and for a single impurity, this equation reduces to:
\begin{equation}
\dot{\hat{e}}_2(q) = \hat{e}_2(q) \left(  - \mu_2 + \mathcal{G}_q \delta \mu  \right) +  ( 1 + \mathcal{G}_q ) \mu_2 a^2 e^{pl} \, \label{eq:EoM2}
\end{equation}
with $\delta \mu =  \mu_3 - \mu_2$. 

\subsection{Equilibrium configuration}
To find the equilibrium configuration, $\dot e_2=0$ is set to zero in eq. \eqref{eq:EoM2}. This immediately leads to
\begin{align}
\hat{e}_2(q) &= a^2 \,  e^{pl}_* \frac {1+\mathcal{G}_q} {1+\mathcal{G}_q\,(1-\mu_3/\mu_2)}\,\mathrm{for}\,q\neq 0\, , \label{eq:e2q} \\
\hat{e}_2(0) &= a^2\,  e^{pl}_* + \frac{V\,\Sigma}{2 \mu_2},
\end{align}
where $V$ is the volume of the system and $e^{pl}_*$ must be determined self-consistently by integrating both sides of eq.~\eqref{eq:e2q} and noticing that $\int_q \hat{e}_2(q) dq \propto e_2(0)$ (for explicit
results, see Appendix~\ref{sec:C}). Regarding the $q=0$-mode (derived from eq.~~\eqref{eq:EoM_inc_real}), i.e., the
extensive total strain $\hat{e}_2(0)$, one may remark that it will remain constant in a strain-controlled protocol, in which
the macroscopic stress $\Sigma$ will thus decrease by $2\mu_2 a^2\,  e^{pl}_* /V$, whereas in a stress-controlled protocol $\hat{e}_2(0)$ will increase by $a^2\, e^{pl}_*$ because of plasticity. For simplicity, the formulae below are given in the case
of a strain-controlled protocol with $\hat{e}_2(0)=0$.

In real space, the Fourier expression of eq.~\eqref{eq:e2q} translates to
\begin{equation}
e_2(r, \theta) = \frac{ C }{r^2}  \frac{\delta + \cos(4\theta)}{(1 + \delta \cos(4\theta))^2}  +  C_0 \delta(r) \,, \label{eq:e2r}
\end{equation}
where the anisotropy parameter $\delta  = \delta \mu/(\mu_3 + \mu_2)$ was introduced below eq.~\eqref{eq:u}, 
and the constants $C_0$ and $C$ are given in Appendix~\ref{app:Fourier} (eq.~\eqref{eq:CC0}), along with details of the derivation.
Finally, we can integrate $e_2(r, \theta)$ to obtain the noteworthy formula of eq.~\ref{eq:u} for the displacement field $u(r,\theta)$, as detailed in Appendix~\ref{sec:ur}. 

Let us mention that these calculations (for an incompressible material) can be generalised to the compressible regime ($B<\infty$, see Appendix~\ref{sec:compressible}) by substituting 
$\mathcal{G}_q$ in eq.~\eqref{eq:e2q} with
\begin{equation}\label{eq:kernelB}
\mathcal{G}_{q}^{B<\infty} = \frac{B}{B + \mu_3} \mathcal{G}_q \,.
\end{equation}

More general expressions giving the elastic field generated by an \emph{extended} impurity can be found in the literature on the mechanics of anisotropic solids\cite{yang1976generalized,kinoshita1971elastic,dunn1997inclusions},
but, being more general, they are also (much) more complex.


\subsection{Connection with elasto--plastic models}\label{sec:EPM}
How do the foregoing results compare with the propagator and rules implemented in elasto-plastic models? To address this question,
we set $a=1$ and focus on the isotropic ($\mu_2 = \mu_3$) and incompressible ($B=+\infty$) case. The elastic strain $\hat{e}_2^{el}(q) := \hat{e}_2  - e^{pl}_* $
then boils down to $e_*^{pl} \mathcal{G}_q$ by virtue of eq.~\eqref{eq:e2q},
where the quadrupolar elastic propagator $\mathcal{G}_q=\frac{-(q_1^2 - q_2^2)^2}{q^4}$ used in EPM is made apparent.

This quadrupolar propagator was derived by Picard and co-workers~\cite{picard04elastic,nicolas2014universal} using a quite different approach, by writing
mechanical equilibrium ($\nabla \cdot \boldsymbol\sigma=0$, where $\boldsymbol\sigma$ is the stress tensor) in an incompressible medium with a plastic eigenstrain $e^{pl}_*$ at the origin. It can also be regarded as the point-wise limit of an Eshelby inclusion \cite{eshelby1957determination}, \emph{i.e.}, a circular inclusion that would spontaneously deform into an ellipse in free space. In real space, the propagator $\mathcal{G}_q$ becomes 
\begin{equation}
\mathcal{G}(r,\theta) = -\mathrm{cos}(4\theta)/\pi r^2 - \delta(r)/ 2 \label{eq:prop_EPM}
\end{equation}
 in polar coordinates [see eq.~\eqref{eq:EshelbyFourier}]. The elastic strain $e_2^{el}(r=0)$ at the origin is thus depressed by $e^{pl}_*/2$ to mitigate the local surge of the plastic strain $e^{pl}_*$, and a quadrupolar halo surrounds the plastic event.

Bearing the foregoing considerations in mind, it is easy to understand that the rules of elasto-plastic cellular automata 
\cite{picard2005slow,lin14epl,nicolas2017deformation} ensue from the choice of the piecewise quadratic potential $V_{2,r}$ sketched in Fig.~\ref{fig:V2} (c) and the neglect of the transient dynamics before mechanical equilibration. Indeed, with this choice, while $e_2(r)$ evolves in a continuous region of $V_{2,r}$, the material is locally Hookean, with $\dot{e}_2(r)=\dot{e}_2^{el}(r)$. Upon reaching a discontinuity of $V_{2,r}$, 
$e_2(r)$ falls into another quadratic branch of the potential, which corresponds to a finite jump of the local plastic strain by an amount $e_*^{pl}(r)$. Using
eq.~\ref{eq:prop_EPM}, this local distortion generates a quadrupolar halo of elastic stress and a depression of the local elastic stress by $e_*^{pl}(r)/2$, viz.,
\begin{align}  
\sigma(r) & \to \sigma(r) - \mu_2 e^{pl}_*(r) \pm \mu_2  e^{pl}_*(r)/V \, \\
 \sigma(r') & \to \sigma(r') + 2 \mu_2 e^{pl}_*(r) \mathcal{G}(r^\prime - r)  \pm \mu_2  e^{pl}_*(r)/V  \label{eq:CA2}
\end{align}
where the sign $\pm$ is positive (negative) for the stress-controlled (strain-controlled, respectively) protocol.

\subsection{Soft modes}\label{sec:SPE2}
One of the advantages of the continuous approach under study over discrete elasto--plastic automata is that, instead of describing a
sequence of mechanical equilibria, it contains the whole dynamics. This, in particular, gives access to the soft modes, whose study was so far
restricted to atomistic simulations \cite{tanguy10mode,manning11soft,charbonneau16universal,yang2017correlations} and, to a lesser extent, 
experiments in which the fluctuations in the particle positions can be imaged \cite{henkes2012extracting}.

In order to perform a linear stability analysis, we assume that 
\begin{equation}
W(e_2) = -  \mu_0 e_2^2 + O(e_2^3)  \,,\, \mu_0 > 0
\end{equation} 
near the equilibrium position $e_2 = 0$, see Fig.~\ref{fig:V2} (a). Decomposing the dynamics of $\hat{e}_2(q,t)$ into a linear sum of Laplace modes, viz., $\hat{e}_2(q,t)= \int_0^{\infty} e^{\omega t} \tilde{e}_2(q,\omega) d\omega$ (so that $\Re(\omega) > 0$ refers to \textit{unstable} modes,
 contrary to the convention of Ref.~\cite{charbonneau16universal,cao2017localisation}), and inserting into eq.~\ref{eq:EoM_inc_real}, we arrive at the following equation:
\begin{equation}
   \left[ \mu_2 - \delta \mu \mathcal{G}_q + \omega \right] \tilde{e}_2(q, \omega)
  =   \mu_0 (1 + \mathcal{G}_q )  a^2  e_2^o(\omega)\, ,\label{eq:softmodegen}
\end{equation}
where $e_2^o(\omega)$ is a shorthand for $\tilde{e}_2(r=0, \omega)$. (For inertial systems,
an additional term $2 \rho \omega^2 q^{-2}$ is present between the brackets on the left-hand side, as detailed in 
Appendix~\ref{app:underdamped}). This equation
determines the shape of the dynamical modes. Indeed, if $e_2^o(\omega)=0$, then $\tilde{e}_2(q, \omega)$ must vanish at
all wavenumbers, except those which cancel the prefactor on the left-hand side of eq.~\ref{eq:softmodegen}.
Bearing in mind that $\mathcal{G}_q \in [-1,0]$, this cancellation is possible if and only if $\omega \in [-\mu_3,-\mu_2]$; there
is thus a continuous range of admissible growth rates $\omega$.
 On the other hand, if $e_2^o(\omega) \neq 0$,  integrating $\tilde{e}_2(q, \omega)$ over $q$
from eq.~\ref{eq:softmodegen} and noticing that $\int_q \tilde{e}_2(q,\omega) \propto e_2^o(\omega)$ 
gives the following closure relation for $\omega$:
 $$ e_2^o(\omega) = e_2^o(\omega) \frac{\mu _0}{\mu _2+\omega } \mathcal{I}\left( \frac{\mu _2-\mu _3}{\mu _2+\omega } \right) \, $$
where the integral function $\mathcal{I}(x)$ is made explicit in eq.~\eqref{eq:integral} of the Appendix. The closure relation is only satisfied for $\omega = \omega_\star$, with
\begin{equation}
\omega_\star = -\mu_2+\frac{\mu _0^2}{2 \mu _0 + \mu _3 - \mu _2} > - \mu_2 \,. \label{eq:omega}
\end{equation}
Since $\omega_\star$ is the maximal growth rate, it is associated with the most unstable mode. Following the same steps as before
to compute this eigenmode from eq.~\eqref{eq:softmodegen}, we find that it derives from the following displacement field:
\begin{align}\label{eq:delta'}
u_{r,\star} \propto \frac{1}{r}  \frac{\cos (2\theta)}{1 + \delta_\star \cos (4\theta) }  \,, u_{\theta,\star} = 0 \,. 
\end{align}
in polar coordinates. Here $ \delta_\star = \frac{\mu _3-\mu _2}{\mu _2+\mu _3+2 \omega_\star }$
quantifies the anisotropy of the elastic medium, as the parameter $\delta$ defined below eq.~\eqref{eq:u}.

Another consequence of eq.~\eqref{eq:delta'} concerns the situations of marginal stability, when $\omega_\star \to 0$, or $\mu_0 \to \mu_2 + \sqrt{\mu_2 \mu_3}$ by eq.~\eqref{eq:omega}.
These situations are on no account mathematical curiosities, but occur whenever a plastic instability is about to 
take place during the \textit{quasi-static} deformation of disordered solids; this instability is triggered by the marginally stable soft mode~\cite{cao2017localisation}. In this case ($\omega_\star \to 0$), the two anisotropy parameters, $\delta$ and
$\delta_\star$, converge and the soft mode (given by eq.~\ref{eq:delta'}) coincides with the equilibrium configuration 
that will be reached after the development of this plastic instability, up to a proportionality coefficient. Besides, this mode
is not affected by the presence of inertia, which becomes negligible for $\omega \to 0$.
Therefore, in quasi-static protocols, we can identify the marginal soft mode and the equilibrium configuration.

\begin{figure}[t]
	\begin{center}
	\includegraphics[width=.9\columnwidth]{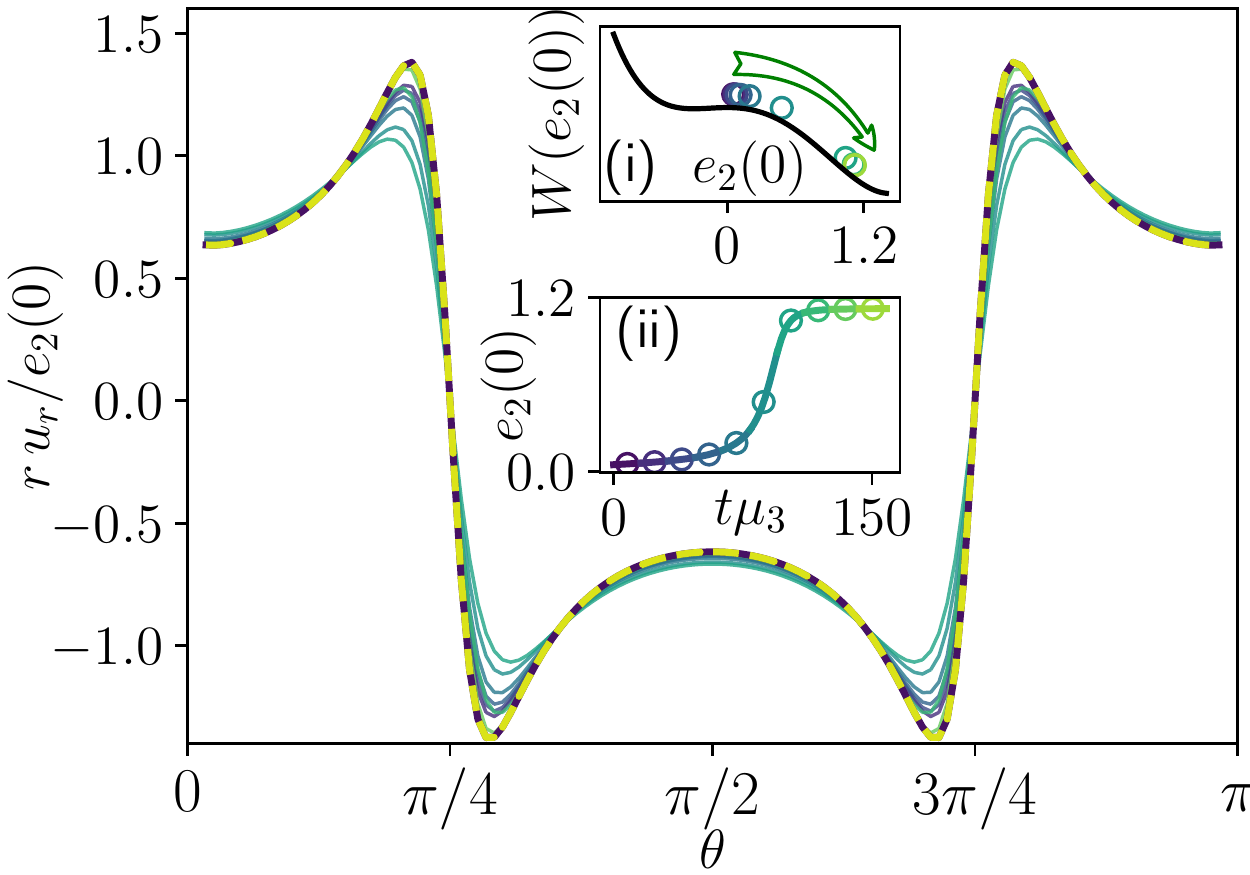}
	\end{center}
	\caption{Time evolution of the displacement field $u(r,\theta)$ in response to a single plastic event for a highly anisotropic incompressible material
at fixed macroscopic strain $\overline{e_2}=0$, with $\mu_2 = 1/19$ and $\mu_3 = 1$ (so that $\delta = 0.9$, as in Fig.~\ref{fig:displacement_th}, (c,f)). We chose an anharmonic potential $W(e_2) = -\mu_0 e_2^2 - \frac32 e_2^3 + e_2^4$, where $\mu_0 = \mu_2 + \sqrt{\mu_2 \mu_3}$ so that $e_2 \equiv 0$ is a marginally stable equilibrium. The initial condition is a small number times the soft mode $\hat{e}_2(q) = 0.04 a^2 (1+\mathcal{G}_q)/( \mu_2 + (\mu_2 - \mu_3) \mathcal{G}_q )$  and the system is evolved according to eq.~\eqref{eq:EoM2} for a duration $T \approx 150$; the evolutions of $e_2(0,t)$ with time and of 
$W$ as a function of $e_2(0)$ are plotted in insets (ii) and (i), respectively.  The main plot shows snapshots of the normalised displacement field at $t = T/10, 2T/10, \dots, T$, computed from $\hat{e}_2$ using the methods of Appendices~\ref{sec:ur} and \ref{app:Fourier}. The colors represent time in the way indicated by the insets. The final configuration is plotted with dashed line since it overlaps with the initial one.} \label{fig:evolution}
\end{figure}

\subsection{Transient dynamics}\label{sec:SPE3}
We close this section by discussing the transient dynamics during the plastic event, i.e., we wonder how the marginal soft mode
gradually unfurls and fades into the new equilibrium state, which is of similar shape. In the special case of isotropic elasticity, i.e., $\mu_2 = \mu_3$, one can check, using eq.~\ref{eq:EoM2}, that the shape is conserved
during the \textit{whole} dynamics, viz.,  $\hat{e}_2(q,t) = e^{pl}(t) a^2  (1 +  \mathcal{G}_q)$, even if the instability
originated in an excited mode ($\omega_\star<0$) and not in the marginal soft mode.
Only the normalisation constant evolves (in the strain-controlled protocol with $\gamma = 0$):
\begin{align*}
  \frac{\dif e^{pl}}{\dif t} =  \mu_2 e^{pl} -  \frac12 W'\left(\frac{e^{pl}}{2}\right) \,.
\end{align*}

The anisotropic case, $\mu_2 \neq \mu_3$, is more involved and requires to solve the equation of motion eq.~\eqref{eq:EoM2} numerically, which can be
done efficiently by noticing that $\hat{e}_2(q)$ depends only on $q_2/q_1$. Perturbing a marginally stable equilibrium configuration $e^0_2$
along a soft mode and letting the system relax at fixed macroscopic strain $\gamma = 0$, we observe that the transient between the initial and final stages is short-lived and only weakly deviates from the equilibrium shape given by eq.~(\ref{eq:u}) (up to normalisation), as shown in Fig.~\ref{fig:evolution}. In particular, in the strongly anisotropic case $\mu_2 \ll \mu_3$,  the equilibrium `fracture'-like profile is apparent during the whole evolution and clearly distinct from the quadrupolar shape found in an isotropic medium. In summary,  the propagators of eq.~\eqref{eq:u} 
are robust signatures of the elastic medium anisotropy, even during the transient dynamics.

\section{Response of a disordered medium with many impurities}\label{sec:denny}
The previous section has shed light on the response of a (possibly anisotropic) elastic medium to a single plastic impurity, with $W_r \propto \delta_{2D}(r)$, and unveiled a
continuous family of elastic propagators. Now, we extend the study to fully disordered media, with many plastic impurities (generic $W_r$).

In this case, the problem becomes analytically intractable, but deserves to be investigated on account of its relevance for collective effects in disordered solids under shear, including avalanches of rearrangements.  A recent analysis of the yielding
behaviour of these materials close to the critical point was notably proposed on the basis of a very similar model by Jagla~\cite{jagla2017non}. Incidentally, even the numerical
study of these models presents difficulties. In particular, the fluctuating sign of the propagator $\mathcal{G}_q$ imposes to carefully follow the order
of the rearrangements. This is in stark contrast with the related, but distinct \cite{lin2014scaling,lin14epl}, problem of elastic line depinning, where the propagator is non-negative (hence the existence of Middleton theorems~\cite{MiddletonPRL}) and efficient algorithms can be devised \cite{werner02rough}.

To proceed, we will inspect the low-frequency excitations of an equilibrium configuration in this generic case by numerically computing
the eigenvectors of the so called dynamical matrix, under the assumption of spatially uncorrelated plastic disorder.


\subsection{Dynamical matrix}\label{sec:matrix}
To start with, we write the general equation of motion (\ref{eq:EoM_inc_real}) in terms of the properly rescaled
\footnote{Using the properly rescaled variable $e$ allows us to obtain a Hermitian dynamical matrix, whereas
a non--Hermitian one is obtained if one considers the variable $e_2$. This is reminiscent of the symmetrization transform from the Fokker--Planck to
the  Schroedinger equation.} strain variable  $e= \mathcal{F}^{-1} e_2$ introduced in eq.~(\ref{eq:rescaled_strain}), with $ \mathcal{F}= Q_3 / q^2$,
\begin{equation}
\dot{e}  =  -\mu_2 e -  \frac{1}{2} \mathcal{F} W_r^\prime\left( \mathcal{F} e \right)   + \frac{\Sigma}{2\mathcal{F}} + (1 - \mathcal{F}^2)  (\mu_2-\mu_3) e \, , \label{eq:edot_overdamped}
\end{equation}
where we have used $\mathcal{F}^2 = 1+ \mathcal{G}_q$.
A small deviation $\delta e$ away from an equilibrium configuration $e^{(0)}$ thus decays as
 $\delta \dot{e}(r) = -\sum_{r^\prime} \mathcal{M}_{rr^\prime} \delta e(r^\prime)$ in discrete space, where the 
dynamical matrix $\mathcal{M}_{rr^\prime}$ reads
\begin{align}
\mathcal{M}_{rr^\prime} = \mu_2 \delta_{rr^\prime}  -  \mathcal{F}  \mathcal{D}_{rr^\prime} \mathcal{F} + (\delta_{rr'} - \mathcal{F}^2)  (\mu_2-\mu_3)& \,  \nonumber \\
\text{with } \mathcal{D}_{rr'} =  -\frac{1}{2} \delta_{rr'} W''_r\left(\mathcal{F} e^{(0)}(r)\right)  \,.& \label{eq:D}
\end{align}
From now on, we focus on the case $\mu_2 = \mu_3$. Then,
\begin{equation}
\mathcal{M}_{rr^\prime} =  \mu_2 \delta_{rr^\prime}  -  \mathcal{F}  \mathcal{D}_{rr^\prime} \mathcal{F}  \label{eq:Mdef}
\end{equation}
is the sum of the scalar matrix $\mu_2 \delta_{rr^\prime}$ and a matrix product between  $\mathcal{F}$ (diagonal in  Fourier space) and $\mathcal{D}_{rr^\prime}$ (diagonal in real space). Accordingly, rescaling the plastic disorder strength as  $\mathcal{D}_{rr^\prime}  \leadsto k \mathcal{D}_{rr^\prime} $ has no effect on the excitation modes, bar an affine transformation of the eigenvalues $\omega \leadsto k(\omega +  \mu_2) -  \mu_2$.
 Once again, the contrast with respect to the equation describing the depinning of an elastic interface should be noted. In that case, the dynamical matrix is a \textit{sum} of a propagator $\mathcal{G}^{\mathrm{(d)}}$ accounting for the elastic couplings within the interface (usually a fractional Laplacian that is diagonal in Fourier space) and a disorder matrix $\mathcal{D}^{\mathrm{(d)}}_{rr^\prime} := \delta_{rr'} W_r^{\mathrm{(d)}\prime\prime}\left(u^{(0)}(r)\right)$ obtained by deriving the disorder potential $W_r^{\mathrm{(d)}}$ with respect to the local {displacement} $u(r)$. The disorder strength can, and does, affect the shape of the eigenmodes, in particular, their localisation length~\cite{cao2017localisation}. 


\subsection{Random approximation}\label{sec:anderson}

\begin{figure}[t]
\begin{center}
	\includegraphics[width=.9\columnwidth]{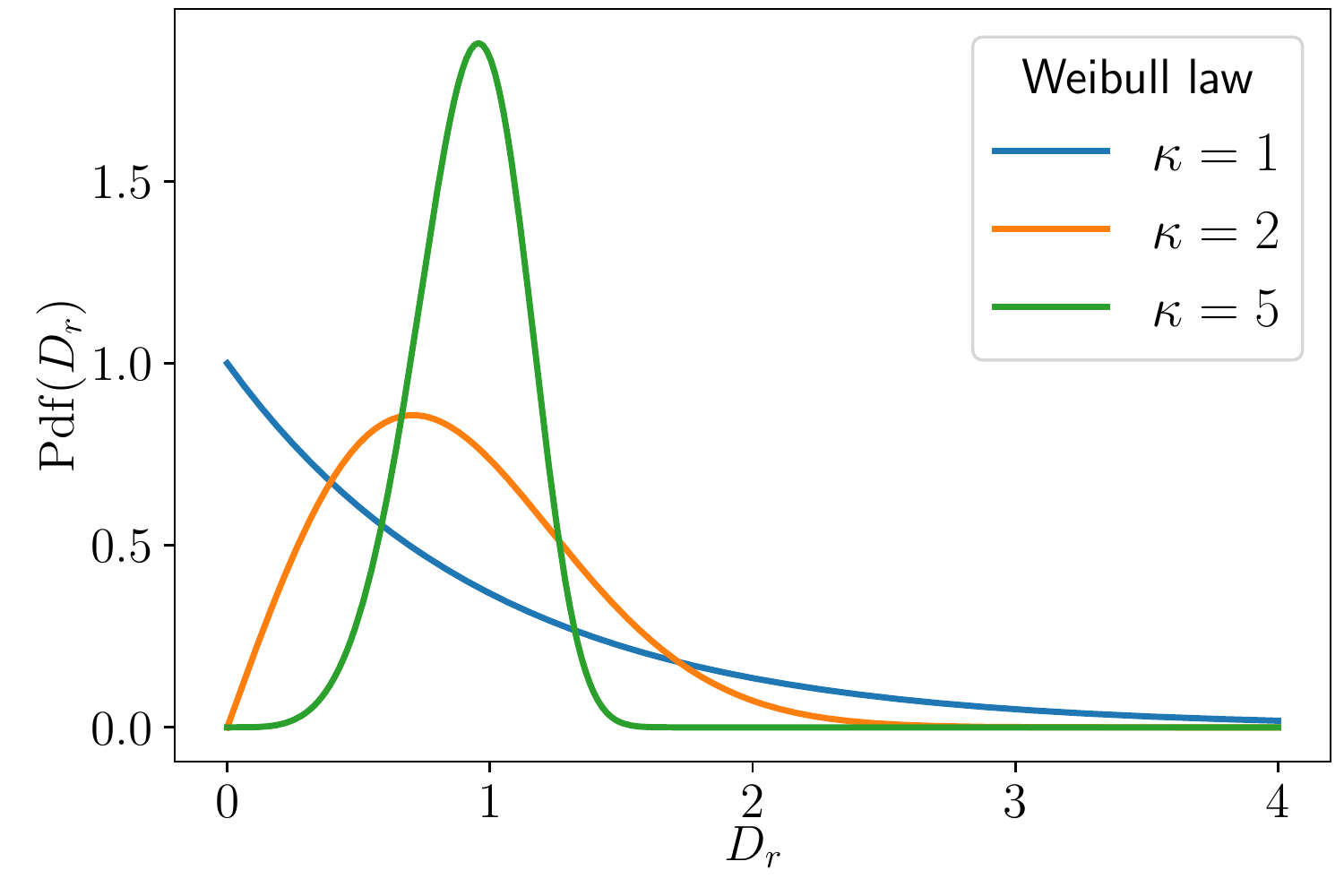}
\end{center}
	\caption{The Probability density function of the Weibull law $\text{Pdf}(D_r) = \kappa D_r^{\kappa-1} e^{-D_r^\kappa}$ for parameters $\kappa = 1, 2,5$.  }\label{fig:weibull}
\end{figure}

\begin{figure*}
	\center
	\includegraphics[width=0.85\textwidth]{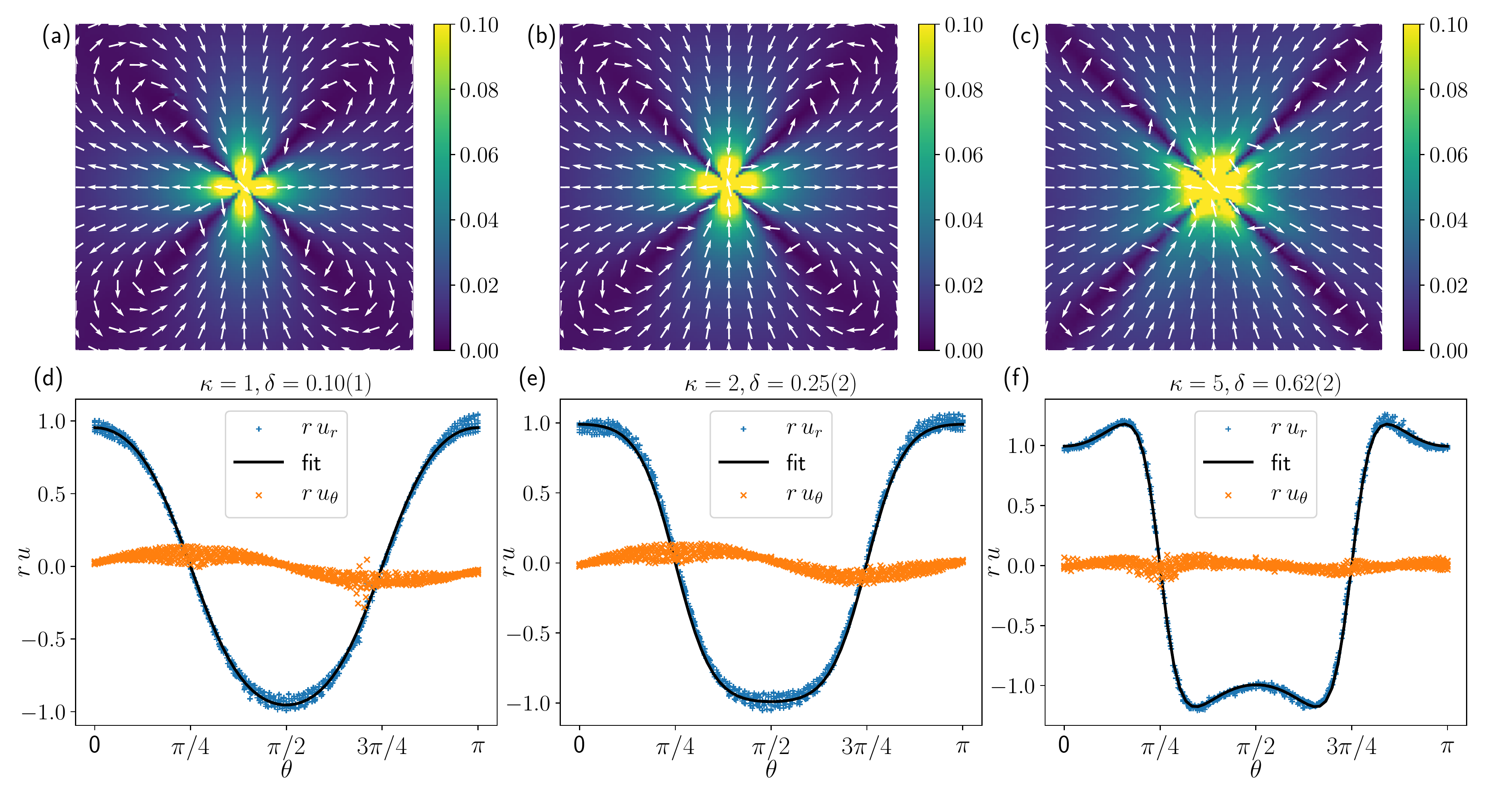}
	\caption{Displacement fields corresponding to the softest modes of the dynamical matrix given in eq.~\eqref{eq:Mdef}, under the random approximation, with Weibull parameters $\kappa = 1, 2, 5$ (see Fig.~\ref{fig:weibull}). For easier comparison with Fig.~\ref{fig:displacement_th},
		we have normalised the plotted fields in the same way as Fig.~\ref{fig:displacement_th}. We centered them around the peak values, and only show a zoom on a square of side $L/2$ (where $L = 256$ is the mesh size) around the peak to get rid of boundary effects.
		The bottom row shows the radial and azimuthal displacements at distance $r \in [L/10, L/5]$ from
		the origin. The solid lines are fits to the Ansatz eq.~\eqref{eq:u} (or equivalently eq.~\ref{eq:delta'}), where the anisotropy parameter $\delta$ is adjusted.} \label{fig:fits}
\end{figure*}

In light of the foregoing observation that the dynamical matrix has the same eigenvectors as $\mathcal{F}  \mathcal{D}_{rr^\prime} \mathcal{F}$, where
 $\mathcal{D}_{rr'}$ is given by eq.~(\ref{eq:D}), 
we are interested in gaining insight into $D_r := - W''_r\left(\mathcal{F} e^{(0)}(r)\right) / 2$.
A priori, it will depend on the specific equilibrium configuration under study and the spatial correlations of $W_r$. However,
the spatial correlations of the disorder will be discarded here, which will spare us the meticulous search of equilibria. 
Albeit uncontrolled, this approximation was recently found to preserve key properties of the eigenmodes at the depinning transition \cite{cao2017localisation}. Indeed, over an ensemble of equilibria, the values taken by $D_r$ at different $r$ are only weakly correlated
and they will be considered random. More precisely, we will handle $D_r$  as independent random variables drawn from a Weibull distribution 
of parameter $\kappa$, which means that $\left(D_r \right)^\kappa$ is exponentially distributed (see Fig.~\ref{fig:weibull}). This enforces $D_r\geqslant 0$ (because plasticity tends to soften the material). The choice of a Weibull distribution is arbitrary, but it will prove convenient,
in that it allows us to tune the dispersion of plastic disorder via the parameter $\kappa$. Considering the limit-cases,  when $\kappa \leq 1$, 
the distribution of $D_r$ is peaked at $D_r = 0$ and, apart from large
 outliers, $D_r \sim 0$.
To the contrary, when $\kappa \gg 1$, a peak around $1$ emerges and the whole distribution concentrates in this peak.

These distributions of $D_r$, combined  with the formula of eq.~\ref{eq:Mdef}, are used to populate random matrices $\mathcal{M}_{rr^\prime}$ (of size $256^2 \times 256^2$, \emph{i.e.}, $r = (x,y), x,y = -128,\dots,127$ and similarly for $r^\prime$). The lowest-frequency eigenvector (softest mode) of each matrix is found numerically by the power/Lancsoz iteration method and is integrated via eq.~\ref{eq:integration} to get the associated displacement field.  Some representative displacement fields for different values of $\kappa$ are plotted in Fig.~\ref{fig:fits}, with the peak value shifted to the origin, for easier comparison with Fig.~\ref{fig:displacement_th}. 
 
Two major conclusions can be drawn from the observation of these plots (among many similar plots). First, irrespective of $\kappa$, we always find 
`soft' modes that are localised, with displacements $u$ and strains $e_2$ 
clearly peaked at an individual site. In other words, no collective pinning (with a peak region spread over many sites) is seen.  At a distance $r$ away from the peak, the radial displacement $u_r$ and the strain
 $e_2$ decay as a power laws $\propto 1/r$ and  $1/r^2$, respectively, while the azimuthal component $u_\theta$ remains very small compared to $u_r$. Secondly, the overall shapes of the displacement fields are strongly reminiscent
 of the elastic propagators derived in Section~\ref{sec:single_impurity} for the single-impurity problem and range from the standard quadrupolar propagator (for small $\kappa$) to the fracture-like propagator (for large $\kappa$). In fact, the theoretical expressions of eq.~(\ref{eq:delta'})
 turn out to fit the numerical displacements quite well, if one is given the freedom to adjust the anisotropy parameter $\delta_\star$
 of the elastic medium, although here $\mu_2=\mu_3$.
 
How can we rationalise these findings? If all sites in the system were decoupled, the softest mode would just be a local excitation of the softest site 
(where $D_r$ is maximal). But, because of the elastic embedding of the site (and, more formally, the St Venant constraint), this excitation
is coupled with an elastic deformation of the surroundings, as though it were an impurity. For small $\kappa$, the disorder $D_r$ is close to zero
on most sites (but not on the softest one, of course), so the medium is virtually a perfect isotropic solid, hence the standard quadrupolar
shape of the mode, that is well captured by an anisotropy parameter $\delta_\star \to 0$. On the other hand, for larger $\kappa$, most sites display 
considerable plasticity-induced softening, with $D_r \approx 1$ (while $D_r>1$ on the softest site). This global softening along $e_2$ is tantamount to 
a lowering of the shear modulus $\mu_2$. Indeed, the dynamical matrix of eq.~(\ref{eq:D}) is not altered by simultaneously reducing $D_r$ by its spatial average
$\overline{D_r}$ and lowering $\mu_2$ to $\mu_2-\overline{D_r}$. Therefore, the impurity is effectively coupled with an anisotropic elastic medium
characterised by a finite $\delta_\star$ and, in the limit of large anisotropy ($\kappa \gg 1$), a fracture-like propagator can emerge, with $\delta_\star \to 1$. Still, it is noteworthy that, even in this regime where the $D_r$ are narrowly distributed around 1, the observed fields remain
dominated by single impurities.

\section{Discussion}

In this work, we have studied an intermediate class of models describing the plastic deformation of disordered solids, that
operate at a coarse-grained scale similar to that of elasto--plastic models (EPM) while still describing the non-linear elastic properties
of atomistic systems. The models considered focus on the (continuous) strain fields
in the material and incorporate them into a Ginzburg-Landau free energy which combines a purely elastic part and a plastic disorder potential.
We investigated the equilibrium configurations, the soft modes and the equilibration dynamics associated with this free energy. 
The last two aspects are lost in EPM, which hop between mechanically equilibrated configurations, following rules that we could clarify.
In contrast, the transient dynamics during mechanical equilibration are present in the continuous models under study, whose
global relaxation thus depends on the deformation rate.\footnote{Nevertheless, the possibility to relax to different energy basins \emph{locally}, depending on the deformation rate, is not taken into account here.}

When plastic disorder is spatially confined in a single impurity, we were able to derive analytically the soft mode and the equilibrium configuration
of the system, which coincide (up to a rescaling factor). Our first important result is that the quadrupolar propagator routinely
used in EPM is not
ubiquitous. Indeed, in the presence of strong elastic anisotropy, we found a new, fracture-like propagator, in which the deformation concentrates
along the easy directions. A continuous family of propagators, obeying a simple formula~\eqref{eq:u} depending on one anisotropy parameter,
interpolates between the quadrupolar propagator and the fracture-like one.

Remarkably, these single-impurity calculations keep being relevant in fully disordered systems, whose lowest-frequency excitations were numerically
found to localise around the softest (point-like) site, even for weakly dispersed disorder. Moreover, the deformation halo around this soft site
replicates the foregoing family of propagators, as the distribution of plastic disorder is varied. The fracture-like propagator is recovered 
for finite, narrowly distributed plastic disorder. Indeed, the latter softens the material along one shearing direction and thus renders the surroundings of the soft site effectively anisotropic.

Accordingly, one may expect to find the fracture-like propagator whenever extended regions of the material collectively soften
on the brink of failure, without immediately failing.
In the presently studied models, this situation is precluded when the disorder potential is piecewise parabolic with
cusps, but should be possible with any smooth potential. Recently, Jagla showed that these two classes of potentials led to different
critical exponents at the yielding transition for the flow curve \cite{jagla2017non}, but the shape of the propagator in each case and
its possible relevance were not studied.

In atomistic simulations and experiments on glasses, non-standard elastic propagators (in the soft mode or the actual stress redistribution) 
have not been reported either, to the best of our knowledge. Certainly, specific conditions are required to observe collective softening
of the material along \emph{one} direction, such as high mechanical homogeneity in the initial configuration, and they may not be met often. In addition, the noise and fluctuations in the numerical and experimental data, notably due to the  granularity of the material at the microscale, may complicate the distinction of a new propagator, all the more so as
the paradigm resting on Eshelby's solution is overwhelming. Nevertheless, we may tentatively expect to see it in
carefully aged (ultra-stable \cite{berthier2017origin}) glasses just before their dramatic macroscopic failure; 
the incipience of a shear band\cite{nguyen16shearband,tanguy10mode} may
also reflect the presence of collective softening. So we advocate to test fits to the different propagators in future simulations and experiments.

\subsection*{Acknowledgments}
We thank S. Bouzat and A.~B. Kolton for collaboration on related
projects, E. Jagla, T. de Geus, P. Le Doussal, A. Tanguy, S. Cazayus-Claverie, and M. Wyart, for insightful discussions. The authors acknowledge support from 
a Simons Investigatorship, Capital Fund Management Paris and LPTMS (X.C.), a LabEx-ICFP scholarship and CNRS (D.T.),  and an ANR grant ANR-16-CE30-0023-01 THERMOLOC (A.R.).

\textit{Conflict of interest.} There are no conflicts of interest to declare.

\appendix

\section{Compressible material}\label{sec:compressible}
In this Appendix, the results derived in the main text for the incompressible case are extended to compressible systems ($B<\infty$).

To this end, the equation of motion \eqref{eq:motion} is generalised to compressible systems, while one still assumes that the
viscosities acting on the different strain variables are equal, $\eta=1$ . Extremising the free energy of eq.~\eqref{Ftot}
with respect to the Lagrange multipliers leads to $\sum_j Q_j \dot{\hat{e}}_j(q) = 0$, for $q \neq 0$. The equation of motion
then straightforwardly generalises to
\begin{equation}\label{eq:motiongen}
\dot{\hat{e}}_j(q) = -\frac{\delta F}{\delta \hat{e}_j(q)} - Q_j  \lambda  \,,\, \lambda = -\frac{\sum_{k=1}^3 Q_k \frac{\delta F}{\delta \hat{e}_j(q)}  }{\sum_{k=1}^3 Q_k^2 }  \,.
\end{equation}
for $j = 1, 2, 3$ and $q\neq0$.  For a single impurity, the above equation can be recast into a matrix form, in terms of 
the 3-vector $\mathbf{e}(q) =\begin{bmatrix}
{\hat{e}}_1(q) & {\hat{e}}_2(q) & {\hat{e}}_3(q) 
\end{bmatrix}$:
\begin{align}
& \dot{\mathbf{e}}  = -\mathbf{P} \left( \mathbf{M} \mathbf{e} +  \mathbf{v} \right)  \label{eq:vectorform}
\end{align}
where  $ \mathbf{v}$ is a $3$-vector, and $\mathbf{P}$ and $\mathbf{M}$ are $3\times3$ matrices defined as follows:
\begin{align}
& \mathbf{P}_{jk} = \delta_{j,k} - \frac{ Q_j Q_k}{\sum_{\ell=1}^3 Q_\ell^2 }\,,\, \mathbf{M}_{jk} = \delta_{j,k} \mu_k \,,\, 
\mu_1 := B\,, \nonumber \\
& \mathbf{v}  = \begin{bmatrix}
0 & \mu_2 a^2 e^{pl} & 0 
\end{bmatrix}  \,. \nonumber
\end{align}
At equilibrium, $\dot{\mathbf{e}}$ vanishes in eq.~\eqref{eq:vectorform}. The strain modes $\mathbf{e}$ that cancel the
right-hand side of  eq.~\eqref{eq:vectorform} and satisfy the St Venant constraint are
\begin{align}
\hat{e}_1(q) &= -a^2 \,  e^{pl}_* \frac{Q_2}{Q_1} \frac{ \mu_3/(B + \mu_3) }{  {1+\mathcal{G}_q^{B<\infty}\,(1-\mu_3/\mu_2)} }  \nonumber \\
\hat{e}_2(q) &=  a^2 \,  e^{pl}_* \frac {1+\mathcal{G}_q^{B<\infty}} {1+ \mathcal{G}_q^{B<\infty}  \,(1-\mu_3/\mu_2)} \label{eq:e2qgen}  \\
\hat{e}_3(q) &=- a^2 \,  e^{pl}_* \frac{Q_2}{Q_3} \frac{ 1 + \mathcal{G}_q^{B<\infty} - \mu_3/(B + \mu_3) }{  {1+\mathcal{G}_q^{B<\infty}\,(1-\mu_3/\mu_2)} } \nonumber
\end{align}
where $e^{pl}_*$ must be determined self-consistently, as detailed in Appendix~\ref{sec:C}, and
$$ \mathcal{G}_q^{B<\infty} = \frac{B}{B + \mu_3} \mathcal{G}_q \,.$$

\section{Plastic strain in the equilibrium situation}\label{sec:C}
Section~\ref{sec:single_impurity} exposed how the shear softening induced by a single plastic impurity at $r=0$, quantified by $e^{pl}(t=0) = - W'(e_2(0,t=0)) / 2 \mu_2$,
generates an
elastic deformation field $e_2(r,t)$, which deforms the impurity and possibly further softens the material at $r=0$.
Equilibrium is reached when 
$$e^{pl}(t) \to e^{pl}_*= -\frac{1}{2 \mu_2} W'(e_2(0)),$$
where $e_2(0)$ forms part of a mechanically equilibrated strain field $e_2(r)$ (no time dependence), or $\hat{e}_2(q)$ in Fourier space.

The Fourier components $\hat{e}_2(q)$ depend on $e^{pl}$ via eq.~\eqref{eq:e2q} (in the incompressible case). Integrating these components over $q$ to get  $ e_2(0) \propto \int \hat{e}_2(q) dq$
leads to
\begin{equation}
e^{pl}_*= -\frac{1}{2 \mu_2} W'\left(\beta e^{pl}_* + \overline{e_2} \right)  \,. \label{eq:plastic1}
\end{equation}
where $\beta= (1+ \sqrt{\mu_3 / \mu_2})^{-1}$. Note that special attention was paid to the zero-mode $\hat{e}_2(0)$, which
is proportional to the average strain $\overline{e_2}$, because its value depends on the deformation protocol,
as explained in the main text.

These results can be generalised to the compressible case $B<\infty$, where the Fourier components $\hat{e}_2(q)$
obey eq.~\eqref{eq:e2qgen} instead of eq.~\ref{eq:e2q}. It turns out that eq.~\eqref{eq:plastic1} still holds,
provided that one replaces $\beta$ with 
$$\frac{B}{B+\mu _3} \mathcal{I}(x) + \frac{\mu_3}{(B+\mu _3)\sqrt{1-x}}.$$
Here, we have introduced the shorthand $x=\frac{B \left(\mu _2-\mu _3\right)}{\mu _2 \left(B+\mu _3\right)}$
and the following integral (that can be calculated by Cauchy's residue theorem for $x<1$),
\begin{equation}\label{eq:integral}
\mathcal{I}(x) = \int_{0}^{2\pi} \frac{\dif \phi}{2 \pi} \frac{\sin (2\phi)^2  }{1 - x \cos(2\phi)^2} = \frac{1}{\sqrt{1-x}+1}.
\end{equation}

\section{Fourier transforms in polar coordinates \label{app:Fourier}}
The calculations in the main text heavily rely on the (convenient) use of Fourier transforms (F.T.), notably in polar coordinates, with
\begin{equation} (x,y)= (r \cos \theta, r \sin \theta) \,,\, q = (\rho \cos \phi, \rho \sin \phi).
\end{equation}
This Appendix collects some results that are useful to derive the real-space expressions from their F.T.

First of all, for functions with an $1/r^2$ dependence (such as the elastic strain generated by a plastic impurity), the following property applies:
\begin{eqnarray} \cos(n \theta) r^{-2} \stackrel{\text{F.T.}} \longrightarrow 2 \pi \im^n\cos(n \phi)  n^{-1} \,,\, n = 1,2,\dots\,. \label{eq:Fourier}
\end{eqnarray}

\emph{Proof.} To derive this result, we recall the Jacobi--Anger identity:
$$ e^{\im q x} = \sum_{m\in \Z} \im^m e^{\im m (\theta -\phi)} J_m(\rho r) $$
$J_m(y)$ is the Bessel $J$ function. Now let $G(x) = e^{\im n \theta} r^{-2}$, with some $n > 0$, then its Fourier transform is
\begin{align}
& \hat{G}(q)
= \int_{0}^{\infty} r^{-1} \dif r \int_0^{2\pi} \dif \theta G(x) e^{\im k x}\nonumber \\
 = & \int_{0}^{\infty} r^{-1} \dif r \int \dif  \theta  G(\theta) \sum_{m\in \Z}  \im^m  e^{\im m (\phi - \theta)} J_m(\rho r) \nonumber \\
 =& 2 \pi  \im^{n}  e^{\im n \phi} \int_{0}^{\infty} r^{-1} \dif r  J_n( \rho r ) = 2 \pi G_n \im^{n}  e^{\im n \phi} n^{-1} 
\end{align}
where in the last line we used the identity
$$ \int_{0}^{\infty} r^{-1} \dif r  J_n(r) = n^{-1}  \,,\,  n = 1, 2, \dots $$
This means that, for $n> 0$, 
\begin{equation}
e^{\im n \theta} r^{-2} \stackrel{\text{F.T.}} \longrightarrow  2 \pi  \im^{n} n^{-1}  e^{\im n \phi} \,,\,
\end{equation} 
which proves eq.~\eqref{eq:Fourier} by parity considerations $\blacksquare$

Now, we specifically turn to the derivation of the real-space field $e_2(r)$ from its F.T. eq.~\eqref{eq:e2q}
 $$\hat{e}_2(q) \propto \frac {1+\mathcal{G}_q} {1+x\mathcal{G}_q},$$  
where $x=1-\mu_3/\mu_2$ and $\mathcal{G}_q = -\cos(2\phi)^2$ by virtue of eq.~\eqref{eq:Gq}.
We start by writing the following identity (obtained \emph{e.g.} via Cauchy residues theorem)
\begin{equation}
\frac {1+\mathcal{G}_q} {1+x\mathcal{G}_q} = 
K + 2 K' \sum_{n=1}^\infty \cos(4n \phi)  z^{n} \,,\, \label{eq:Cauchy}
\end{equation}
where $K = (\sqrt{1-x}+1)^{-1}$, $K' = -\sqrt{1-x}/x$ and $z =x / (1 +\sqrt{1-x})^2$. 
Then we apply an inverse F.T. to eq.~\eqref{eq:Cauchy} term by term with the help of eq.~\eqref{eq:Fourier} and arrive at
\begin{equation}
e_2(r) \propto
 \frac{2}{\pi} \frac{\sqrt{1-x}}{x-2} \frac{\delta + \cos(4\theta)}{(1 + \delta \cos(4\theta))^2} \frac{1}{r^2} + \frac{\delta_{2D}(r)}{1 + \sqrt{1-x}} \,. \label{eq:inverseF}  
\end{equation}
where  $\delta = \frac{x}{x-2}$.
This result is consistent with eq.~\eqref{eq:e2r}, provided that
\begin{equation} C = -\frac{2 a e_*^{pl}}{ \pi} \frac{\sqrt{\mu_2 \mu_3}}{\mu_2+\mu_3} \,,\, C_0 =  \frac{a^2 e_*^{pl}}{\sqrt{\mu_3/\mu_2}+1}  \,. \label{eq:CC0}\end{equation}

Incidentally, the formula for the isotropic case is recovered for $x = 0$,
\begin{equation} 1+\mathcal{G}_q = \frac{1}{2} - \frac{1}{2}\cos(4\phi) \stackrel{\text{F.T.}^{-1}}{\longrightarrow}  \frac{\delta(r)}{2} - \frac{ \cos(4\theta)} {\pi r^{2}} \,.\label{eq:EshelbyFourier}
 \end{equation}

 \section{Displacement field and strain}\label{sec:ur}
 We explicit the relation between displacement field and strain tensor, for a displacement field which points in the radial direction, and whose amplitude $\Vert u \Vert = F(\theta) / r$. The corresponding strain tensor is as follows [see eq.~\eqref{eq:tensors} and text below]:
 \begin{equation}\label{eq:e2diff}
 e_1 = 0 \,,\, e_2 = - \frac1{2r^2}\frac{\dif \left[\sin (2 \theta ) F(\theta ) \right]} {\dif\theta}\,.
 \end{equation}
 So the displacement is incompressible for any $F$. Comparing to eq.~\eqref{eq:e2r} yields a differential equation for $F(\theta)$, whose general solution is:$$ F(\theta) = -C \frac{\cos(2\theta)}{1+\delta \cos(4\theta)} + c_1 \csc(2\theta) \,,$$
 where $c_1$ is an integral constant which we set to $0$ (since $\csc(2\theta)$ is divergent at $\theta = 0, \pi/2, \pi, 3\pi/2$), leading to eq.~\eqref{eq:u}.
 
 Eq.~\eqref{eq:e2diff} was also used to obtain the displacement field in the numerical simulation of transient dynamics in Section~\ref{sec:SPE3}.

\section{Generalisation to underdamped systems} \label{app:underdamped}
	The main text focuses on the overdamped limit relevant for foams and concentrated emulsions, notably. Here, the
	results are generalised to all damping regimes. Following the approach of \cite{lookman03ferro}, this is achieved by
	complementing the Euler-Lagrange equations of motion with a strain-rate-dependent Rayleigh dissipation term
	$R = \eta \int_r \left[\dot{e}_2^2 + \dot{e}_3^2  \right] = \eta \int_r \dot{e}^2$, where $\eta=1$ is the viscosity and the rescaled strain
	$e$ was introduced in eq.~\eqref{eq:rescaled_strain}. The resulting equations read
	
	\begin{equation}
    \frac{d}{dt} \frac{ \delta \mathcal{L} }{\delta \dot{e}} -  \frac{ \delta \mathcal{L} }{\delta e}  = - \frac{\delta R}{\delta \dot{e}} \,,
     \end{equation}
	where the Lagrangian
	\begin{equation}
	 \mathcal{L}= T- F
	\end{equation}
	is the difference between the kinetic energy  $T=\frac{1}{2}\int_r \rho \dot{u}^2$  and the free energy
     \begin{equation}
     F = \int_r  \left[W_{r}[e_2(r)] +  \mu_2 e_2(r)^2 + \mu_3 e_3(r)^2 \right] \,.
     \end{equation}

     Since displacement is a (non-local) function of the strain field, the kinetic energy $T$ can be expressed in terms of the strain rate, 
     instead of the velocity. Adapting the results of \cite{lookman03ferro} (Sec.~III.A) to an incompressible system, we obtain
     \begin{equation}
     T =  2 \rho \int_q \frac{ |\dot{\hat{e}}|^2 (q) }{q^2} \,.
     \end{equation}
	 
	 The Lagrange-Rayleigh equations then simplify to
     \begin{align}
  2 \rho  \ddot{e} = - q^2 \left( \frac12 \frac{Q_3}{q^2}  \frac{\partial W_{r}}{\partial e_2(r)}   +  \mu_2 e  +  \delta \mu  \frac{Q_2^2}{q^4}  e  +  \dot{e} \right) \,, \label{eq:e}
     \end{align}
      where $\delta \mu = \mu_3 - \mu_2$. Note that the overdamped dynamics of eq.~\eqref{eq:edot_overdamped} are recovered for
      $\rho \to 0$.
      
     Linearising the above equation and writing $\hat{e}(q,t)= \int_0^{\infty} e^{\omega t} \tilde{e}(q,\omega) d\omega$ yields
    \begin{equation}
     2 \rho \omega^2 \tilde{e} =  q^2  \left( \mathcal{F} \mathcal{D} \mathcal{F} + \delta \mu \, \mathcal{G}_q - \mu_2 -  \omega \right) \tilde{e} \,, \label{eq:app_EoM}
     \end{equation}
     where $\mathcal{F} = Q_3 / q^2$, $\mathcal{G}_q=\mathcal{F}^2-1$, and $\mathcal{D} =  -\frac{1}{2} \delta_{rr'} W''_r\left(\mathcal{F} e^{(0)}(r)\right)$ were already defined in Sec.~\ref{sec:denny}. Only in the marginal case $\omega \to 0$ do the resulting eigenmodes coincide with those
      of the overdamped dynamical matrix of eq.~\eqref{eq:D}. Otherwise, the propagation of sound waves affects the transient dynamics. 
      
      The foregoing statement can be made more explicit in the case of a single impurity, i.e., $W_r =a^2 \delta(r) W$. While the equilibrium configuration does not depend on the damping, the transient dynamics do. In particular, the vibrational modes from eq.~\eqref{eq:app_EoM},
      expressed in terms of the original strain variable $e_2$, read
       \begin{equation}
       \tilde{e}_2(q) \propto  \frac{1 +\mathcal{G}_q}{2 \rho \omega^2 q^{-2} + \omega + \mu_2 - \delta \mu \, \mathcal{G}_q},
       \end{equation}
  which matches the (overdamped) eigenmodes of eq.~\eqref{eq:softmodegen} only for the marginally stable mode at $\omega = 0$.

\bibliographystyle{rsc}
\bibliography{yield}
\end{document}